\documentclass[galaxies,review,accept,moreauthors,pdftex,10pt,a4paper]{mdpi} 

\newcommand{\gtrsim}{\ ^{\displaystyle >}_{\displaystyle \sim}\ }
\newcommand{\lesssim}{\ ^{\displaystyle <}_{\displaystyle \sim}\ }

\firstpage{1} 
\pubvolume{6}
\issuenum{1}
\articlenumber{1}
\pubyear{2018}
\copyrightyear{2018}
\history{Received: 25 October 2018; Accepted:21 December 2018; Published: date}
\Title{Halo Concentrations and the Fundamental Plane of Galaxy Clusters}

\Author{Yutaka Fujita $^{1,}$*\orcidA{}, Megan Donahue $^{2}$, Stefano Ettori $^{3,4}$, Keiichi Umetsu $^5$, Elena Rasia $^6$, Massimo Meneghetti $^{3,4}$, Elinor Medezinski $^{7}$, Nobuhiro Okabe $^{8}$ and Marc Postman $^{9}$
}
\AuthorNames{Yutaka Fujita}

\address{%
$^{1}$ \quad Department of Earth and Space Science, Graduate School of Science, Osaka University, Toyonaka, Osaka~560-0043, Japan\\
$^{2}$ \quad Physics and Astronomy Department, 
Michigan State University, East Lansing, MI 48824, USA; donahue@pa.msu.edu\\
$^{3}$ \quad INAF, Osservatorio di Astrofisica e Scienza dello Spazio, via Pietro Gobetti 93/3, 40129 Bologna, Italy; ettori.s@gmail.com (S.E.); massimo.meneghetti@gmail.com (M.M.)\\
$^{4}$ \quad INFN, Sezione di Bologna, viale Berti Pichat 6/2, I-40127 Bologna, Italy\\
$^{5}$ \quad Institute of Astronomy and Astrophysics, Academia
Sinica, P.O. Box 23-141, Taipei 10617, Taiwan; keiichi@asiaa.sinica.edu.tw\\
$^{6}$ \quad INAF, Osservatorio Astronomico di Trieste, via Tiepolo 11,
 I-34131 Trieste, Italy; elena.rasia.oats@gmail.com\\
$^{7}$ \quad Department of Astrophysical Sciences, 
4 Ivy Lane, Princeton, NJ 08544, USA; elinorm@astro.princeton.edu\\
$^{8}$ \quad Department of Physical Science, Hiroshima University, 1-3-1
 Kagamiyama, Higashi-Hiroshima, Hiroshima~739-8526, Japan; okabe@hiroshima-u.ac.jp\\
$^{9}$ \quad Space Telescope Science Institute, 3700 San Martin Drive,
 Baltimore, MD 21208, USA; postman@stsci.edu
}

\corres{Correspondence: fujita@astro-osaka.jp; Tel.: +81-6-6850-5484}

\abstract{According to the standard cold dark matter (CDM) cosmology,
the structure of dark halos including those of galaxy clusters reflects
their mass accretion history. Older clusters tend to be more
concentrated than younger clusters. Their structure, represented by the
characteristic radius $r_s$ and mass $M_s$ of the Navarro--Frenk--White
(NFW) density profile, is related to their formation time. In~this
study, we showed that $r_s$, $M_s$, and the X-ray temperature of the
intracluster medium (ICM), $T_X$, form a thin plane in the space of
$(\log r_s, \log M_s, \log T_X)$. This tight correlation indicates that
the ICM temperature is also determined by the formation time of
individual clusters. Numerical simulations showed that clusters move along
the fundamental plane as they evolve. The plane and the cluster
evolution within the plane could be explained by a similarity solution of
structure formation of the universe. The angle of the plane shows that
clusters have not achieved ``virial equilibrium'' in the sense that
mass/size growth and pressure at the boundaries cannot be ignored. The
distribution of clusters on the plane was related to the intrinsic
scatter in the halo concentration--mass relation, which   originated
from the variety of cluster ages. The well-known mass--temperature
relation of clusters ($M_\Delta\propto T_X^{3/2}$) can be explained by
the fundamental plane and the mass dependence of the halo concentration
without the assumption of virial equilibrium. The fundamental plane could
also be used for calibration of cluster masses.}

%NOTE: Please confirm these keywords.

\begin{document}
%%%%%%%%%%%%%%%%%%%%%%%%%%%%%%%%%%%%%%%%%%
%% Only for the journal Gels: Please place the Experimental Section after the Conclusions

%%%%%%%%%%%%%%%%%%%%%%%%%%%%%%%%%%%%%%%%%%

\section{Introduction}
\label{sec:intro}

Clusters of galaxies are the most massive objects in the Universe. Since
the fraction of baryons in clusters is not much different from the
cosmic mean value, dark matter accounts for most of the mass of clusters
($\sim$84\%) \cite{2011ApJS..192...18K,2018arXiv180706209P}. Thus, the
structure of the clusters is mainly determined by the halos of dark
matter, or the dark halos. Cold dark matter (CDM) cosmology predicts
that more massive halos   form  later. Thus, clusters form after
galaxies do. However, the definition of the formation is not obvious,
because halos are continuously growing through mergers and accretion
from their environments. A current trend may be associating the
formation time with the internal structure of dark halos.

The density distribution of dark halos is well-represented by the
Navarro--Frenk--White (NFW) density profile \cite{nav96a,nav97a}:
\begin{equation}
\label{eq:NFW}
 \rho_{\rm DM}(r) = \frac{\delta_c\rho_c}{(r/r_s)(1+r/r_s)^2}\:,
\end{equation}
where $r$ is the clustercentric radius, $r_s$ is the characteristic
radius, and $\rho_c$ is the critical density of the universe. The
normalization of the profile is given by $\delta_c$. The characteristic
mass $M_s$ is defined as the mass inside $r_s$ and the characteristic
density is written as $\rho_s\equiv 3\: M_s/(4\pi r_s^3)$.
The mass profile of the NFW profile is written as
\begin{equation}
\label{eq:MNFW}
 M(r) = 4\pi\delta_c\rho_c r_s^3
\left[\ln\left(1+\frac{r}{r_s}\right)-\frac{r}{r+r_s}\right]\:.
\end{equation}

{ Another commonly defined characteristic radius of clusters is that
based on the critical density $\rho_c$;} it is represented by
$r_\Delta$, which is the radius of a sphere of mean interior density
$\rho_\Delta\equiv \Delta\rho_c$, where $\Delta$ is the constant. The
mass within $r_\Delta$ is written as
\begin{equation}
\label{eq:MD}
 M_\Delta = \frac{4\pi}{3}\rho_\Delta r_\Delta^3\:.
\end{equation}

The radius when $\Delta=200$ or $r_{200}$ is often called the ``virial
radius''. Since it is generally difficult to observationally study
cluster properties out to $r\sim r_{200}$, $\Delta=500$ is also used as
a representative value. The ratio
\begin{equation}
 \label{eq:cD}
c_\Delta=r_\Delta/r_s
\end{equation}
is called the halo concentration parameter and $c_\Delta>1$ for
$\Delta=200$ and 500 for clusters.

Navarro et al. \cite{nav97a} pointed out that the characteristic
parameters of the NFW profile (e.g., $\rho_s$ and $c_\Delta$) reflect
the density of the background universe when the halo was formed. This
means that, since older halos formed when the density of the universe
was higher, they tend to have larger characteristic densities $\rho_s$
and become more concentrated with larger $c_\Delta$.  This issue has
been addressed in many studies, especially by $N$-body simulations
\cite{nav97a,bul01a,2001ApJ...554..114E,wec02a,zha03a,2006ApJ...646..815S,2007MNRAS.381.1450N,2008MNRAS.387..536G,zha09a,2012MNRAS.423.3018P,lud13a,cor15c,cor15b,mor15b,2018arXiv181008212R}. These
studies have indicated that the halo structure is determined by their
mass-growth history. The inner region ($r\lesssim r_s$)\footnote{More
precisely, the boundary radius between the inner and outer regions is a
few times $r_s$ (e.g. \cite{zha03a}).} of current halos develops in the
early ``fast-rate growth'' phase when the halos grow rapidly through
matter accumulation. Their outer region ($r\gtrsim r_s$) is formed in
the subsequent ``slow-rate growth'' phase in which halos grow slowly
through moderate matter accumulation. During this phase, the inner
region is almost preserved. Thus, halos form ``inside-out''. The
formation time of a halo can be defined as the transition time from the
fast-rate growth phase to the slow-rate growth phase. This shift of the
growth phase is largely associated with the decrease in the average
density of the universe in the $\Lambda$CDM cosmology.  There~are a few
specific definitions of the formation time that well represent the
transition time. One is the time at which the mass of the main
progenitor was equal to the characteristic mass $M_s$ of the halo at its
observed redshift $z_{\rm obs}$ \cite{lud13a,cor15c}. The formation
redshift ($z_f$) corresponding to the formation time should be larger
than $z_{\rm obs}$, or $z_f\geq z_{\rm obs}$. For a given $z_{\rm obs}$,
clusters with a larger $z_f$ have a larger $\rho_s$ and $c_\Delta$.

Moreover, numerical simulations have shown that clusters are dynamically
evolving systems and such evidence is often found in their outskirts. In
fact, the ambient material is continuously falling toward clusters,
which creates ``surfaces'' around clusters. For example, the outskirt
profiles of dark matter halos can become extremely steep over a narrow
range of radius (``splashback radius''). This features in the density
profiles are caused by splashback of collisionless dark matter on its
first apocentric passage after accretion
\cite{2014JCAP...11..019A,2014ApJ...789....1D}. Accretion of collisional
gas toward clusters also creates discontinuities in the form of shock
fronts in their outskirts
\cite{2000ApJ...542..608M,2003ApJ...593..599R}. These discontinuities
mean that clusters are neither isolated nor in an equilibrium state.

In this review, we first introduce the fundamental plane we
discovered, and its implications for structure formation of the
universe. In particular, we show that clusters in general have not
achieved virial equilibrium in contrast with conventional views. Then,
using the fundamental plane, we discuss that a scaling relation
(mass-temperature relation) can be explained without assuming virial
equilibrium. We also show that the fundamental plane can be used for
mass calibration. We assume a spatially-flat $\Lambda$CDM cosmology
with $\Omega_\mathrm{m}=0.27$, $\Omega_\Lambda=0.73$, and the Hubble
constant of $H_0=70$\,km\,s$^{-1}$\,Mpc$^{-1}$.

\section{Fundamental Plane}

The hot intracluster medium (ICM) is distributed in the potential well
of dark halos. Since the X-ray emission from the ICM is proportional to
the square of the density, it mainly comes from the central region of
the cluster where the density is high. Thus, the observed X-ray
temperature $T_X$ represents that of the central region and should
reflect the gravitational potential there. Since the potential is
determined by $r_s$ and $M_s$, we can expect some relation among $T_X$,
$r_s$, and $M_s$.

\begin{figure}[H]
\centering \includegraphics[width=12 cm]{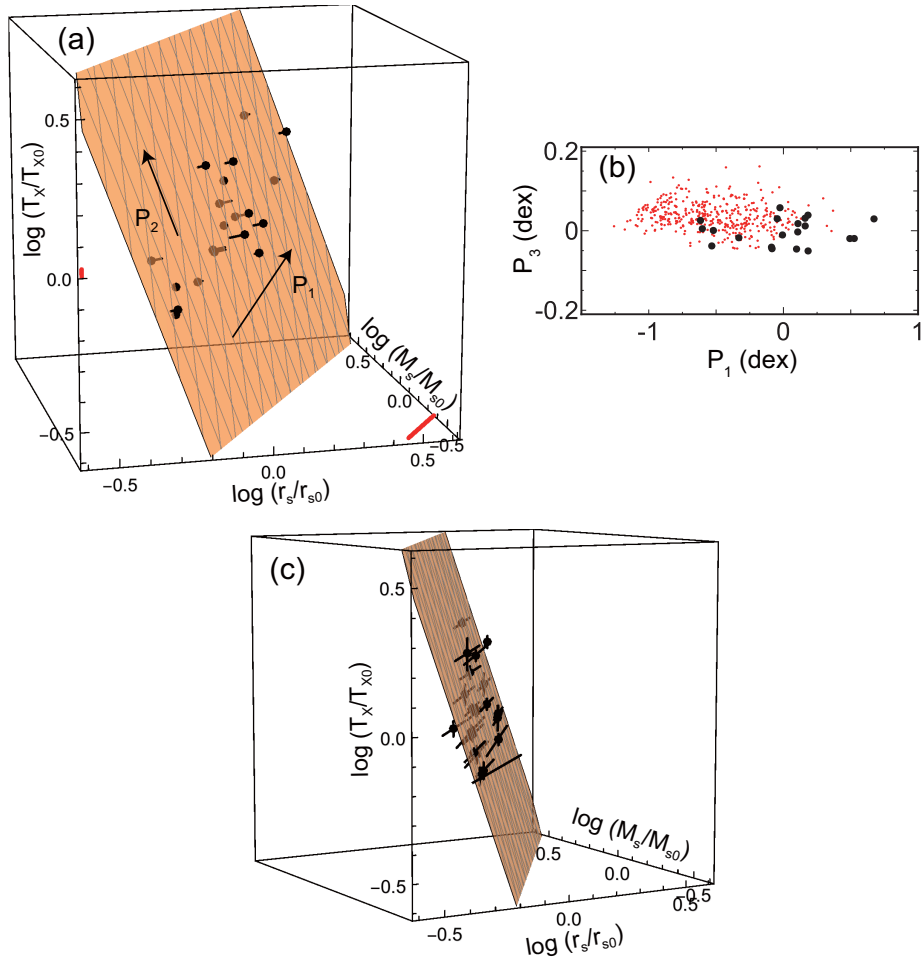}
\caption{(\textbf{a}) Black dots (pin heads) are the CLASH data in the
space of $(\log (r_s/r_{s0})$, $\log (M_s/M_{s0})$, $\log
(T_X/T_{X0}))$, where $r_{s0}=570$~kpc, $M_{s0}=3.8\times 10^{14}\:
M_\odot$, and $T_{X0}=8.2$~keV are the sample geometric averages (log
means) of $r_s$, $M_s$, and $T_X$, respectively. The orange plane is the
best fit of the data. The~orange plane is translucent, and the grayish
points are located below the plane. The lengths of the pins show the
distance to the plane. The red bars show typical $1\sigma$ errors of the
data. The arrow $P_1$ shows the direction to which the data distribution
is most extended, and the arrow $P_2$ is perpendicular to $P_1$ on the
plane. (\textbf{b}) The cross-section of the plane in (\textbf{a}). The
large black dots are the CLASH data, and the small red dots are the
MUSIC results shown in Figure~\ref{fig:FP_MUSIC}.  The latter is
projected on the $P_1$--$P_3$ plane determined for the former.  The
direction $P_3$ is the plane normal. Note that the scales of the
vertical and horizontal axes are different. (\textbf{c}) The same as
(\textbf{a}) but error bars for individual clusters are included. The
viewing angle is changed so that the relation between the error bars and
the plane is easily seen (Figure is reconstructed from Figure~1 of
\cite{fuj18a}).}  \label{fig:FP_CLASH}
\end{figure}

\begin{figure}[H]
\centering \includegraphics[width=6.5 cm]{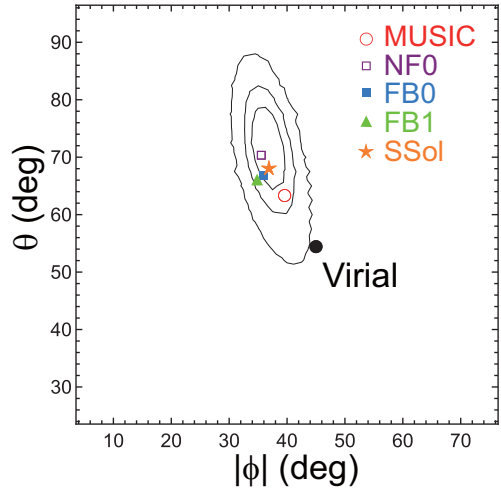} \caption{The
angle of the plane normal $P_3$ in the space of $(\log r_s, \log M_s,
\log T_{\rm X})$; $\theta$ is the angle between $P_3$ and the $\log
T_{\rm X}$ axis, and $\phi$ is the azimuthal angle around the $\log
T_{\rm X}$ axis, measured anti-clockwise from the $\log r_s$ axis. The
contours are for the CLASH observational data showing the 68\%
(1$\sigma$), 90\% and 99\% confidence levels from inside to outside. The
large black dot (Virial) is the prediction of the simplified dimensional
analysis or $T_s\propto M_s/r_s$ and is rejected at the >99\%
level. The directions of the plane normals estimated for the simulation
samples MUSIC, NF0, FB0, and FB1 are shown by the open red circle, the
open purple square, the filled blue square, and the filled green
triangle, respectively. The~prediction of the similarity solution
(Equation~(\ref{eq:FP}) for $n=-2$) is shown by the orange star (SSol)
(Figure is reconstructed from Figure~2 of \cite{fuj18a}).}
\label{fig:P3angle}
\end{figure}

Based on this motivation, Fujita et al. \cite{fuj18a} analyzed 20 massive
  clusters in the Cluster Lensing And Supernova survey with Hubble
(CLASH) observational sample \cite{pos12a,men14a}.  For these clusters,
$r_s$ and $M_s$ had been obtained from the joint analysis \cite{ume16a}
of strong lensing observations with 16-band Hubble Space Telescope
observations \cite{2015ApJ...801...44Z} and weak-lensing observations
mainly with Suprime-Cam on the Subaru Telescope \cite{ume14a}. The X-ray
temperature had been obtained with {\em Chandra}
\cite{pos12a,don14a}. Temperatures are estimated for a cylindrical
volume defined by the projected radii $r=50$--500~kpc to avoid the
influence of cool cores.  Figure~\ref{fig:FP_CLASH}a shows the data
distribution in the $(\log r_s, \log M_s, \log T_X)$ space. As~can be
seen, the data are distributed on a plane. { The plane is described
by $r_s^a M_s^b T_X^c=\mathrm{const.}$, with $a=0.76^{+0.03}_{-0.05}$,
$b=-0.56^{+0.02}_{-0.02}$, and $c=0.32^{+0.10}_{-0.09}$.}
Figure~\ref{fig:FP_CLASH}b shows the cross-section of the plane; the
dispersion of the data around the plane is very small and is only
$0.045^{+0.008}_{-0.007}$ dex (all uncertainties are quoted at the
$1\:\sigma$ confidence level unless otherwise mentioned). { In
Figure~\ref{fig:FP_CLASH}c, error bars for individual clusters are
shown. In the vertical direction ($T_X$), we show the temperature errors
of individual clusters. In the horizontal direction, the errors of $r_s$
and $M_s$ are strongly correlated, and we display them as a single
bar. This means that we draw a bar connecting ($r_s^u$, $M_s^u$) and
($r_s^l$, $M_s^l$) for each cluster, where the superscripts $u$ and $l$
are the upper and the lower limits, respectively. We note that, when we
calculate the plane parameters, we   properly account  for the
correlation for each cluster using the joint posterior probability
distribution of the NFW parameters (mass and concentration)
\cite{fuj18a}. Thus, the actual error is not represented by a single bar
in a precise sense. Figure~\ref{fig:P3angle} shows the direction of the
plane normal $P_3$ in the $(\log r_s, \log M_s, \log T_X)$ space
\cite{fuj18a}. The contours are drawn considering the errors and show
that the direction is inconsistent with the prediction of a simplified
dimensional analysis or $T_X\propto M_s/r_s$.
We note that it is meaningless to discuss cluster distribution in the
space of $(\log r_{200}, \log M_{200}, \log T_X)$, where $r_{200}$,
$M_{200}$, and $T_X$ are their {\it current} values. Although the
clusters form a plane in that space, it is just the obvious relation of
$M_{200}=4\pi \rho_{200} r_{200}^3/3 \propto r_{200}^3$ regardless of
$T_X$ (Section 5.3 in \cite{fuj18a}).}

\begin{figure}[H]
\centering \includegraphics[width=12 cm]{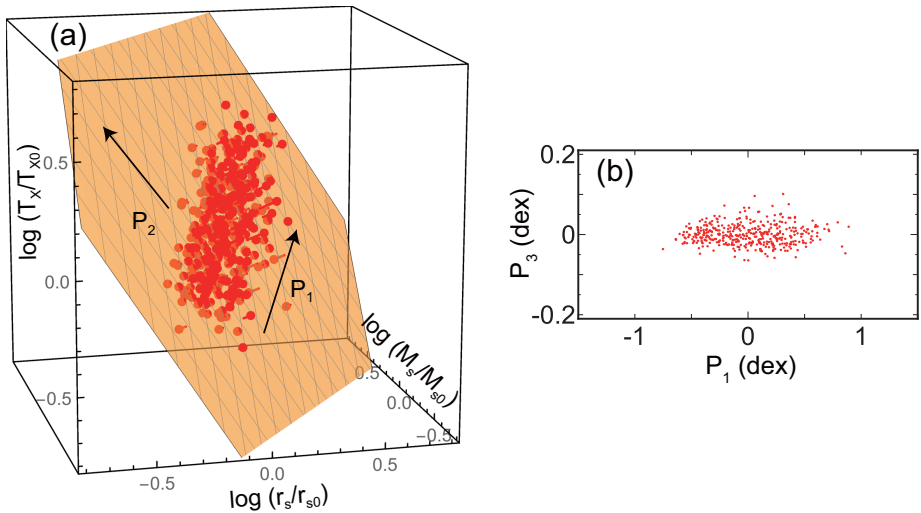} 
%\caption{\hl{The same as
%Figure}~\ref{fig:FP_CLASH} but for the adiabatic MUSIC simulations%NOTE: Please confirm whether this is acceptable or if the full descriptions needs to be re-written (here and other similar figures below)
%($z=0.25$). The axes are normalized by the average parameters of the
%sample ($r_{s0}=414$~kpc, $M_{s0}=1.4\times 10^{14}\: M_\odot$, and
%$T_{X0}=3.7$~keV). (Figure is reconstructed from Figure~3 of
%\cite{fuj18a}).\textcolor{red}{We appreciate full descriptions re-written here instead of `the same as Figure 1' .\\please explain the subfigures a and b in the caption. }}
\caption{(\textbf{a}) Red dots (pin heads) are the results of the
adiabatic MUSIC simulations ($z=0.25$).  The axes are normalized by the
average parameters of the sample ($r_{s0}=414$~kpc, $M_{s0}=1.4\times
10^{14}\: M_\odot$, and $T_{X0}=3.7$~keV). The orange plane is the best
fit of the data. The arrow $P_1$ shows the direction to which the data
distribution is most extended, and the arrow $P_2$ is perpendicular to
$P_1$ on the plane. (\textbf{b}) The cross-section of the plane in
(\textbf{a}). The direction $P_3$ is the plane normal (Figure is
reconstructed from Figure~3 of \cite{fuj18a}).}  \label{fig:FP_MUSIC}

\end{figure}

The ``fundamental plane'' \footnote{Other fundamental planes of clusters
with different combinations of three parameters have also been studied
\cite{1993MNRAS.263L..21S,1998A&A...331..493A,1999ApJ...519L..51F,2002ApJ...581....5V,2004ApJ...600..640L,2006ApJ...640..673O,2009MNRAS.400.1317A}.}%NOTE: Footnote.
has been reproduced by numerical simulations. Figure~\ref{fig:FP_MUSIC}
shows the results of MUSIC N-body/hydrodynamical simulations (see
details in \cite{men14a,fuj18a}). These simulations do not include
radiative cooling or non-gravitational feedback by active galactic
nuclei (AGNs) and supernovae (SNe). { The mass resolutions for the
dark-matter particles that for the gas particles are $m_{\rm
DM}=9.01\times 10^8\: h^{-1}\: M_\odot$ and $m_{\rm SPH}=1.9\times
10^8\: h^{-1}\: M_\odot$, respectively, where the Hubble constant is
written as $H_0=100\: h$\,km\,s$^{-1}$\,Mpc$^{-1}$ and $h=0.7$. The
gravitational softening is set to be 6 $h^{-1}$~kpc for the both gas
and dark-matter particles in high-resolution regions. We chose all  
  402 clusters at $z=0.25$ with $M_{200}>2\times 10^{14}\: h^{-1}\:
M_\odot$ regardless of dynamical state.} In this analysis, we included
the core because these simulations are non-radiative and thus do not
present cool-core features.  The~absolute position of the plane is very
close to that of the CLASH observational data
(Figure~\ref{fig:FP_CLASH}b). Figure~\ref{fig:P3angle} shows that the
plane angle for the MUSIC sample is consistent with the CLASH data at
the 90\% confidence level.

\vspace{-6pt}

\begin{figure}[H]
\centering \includegraphics[width=12 cm]{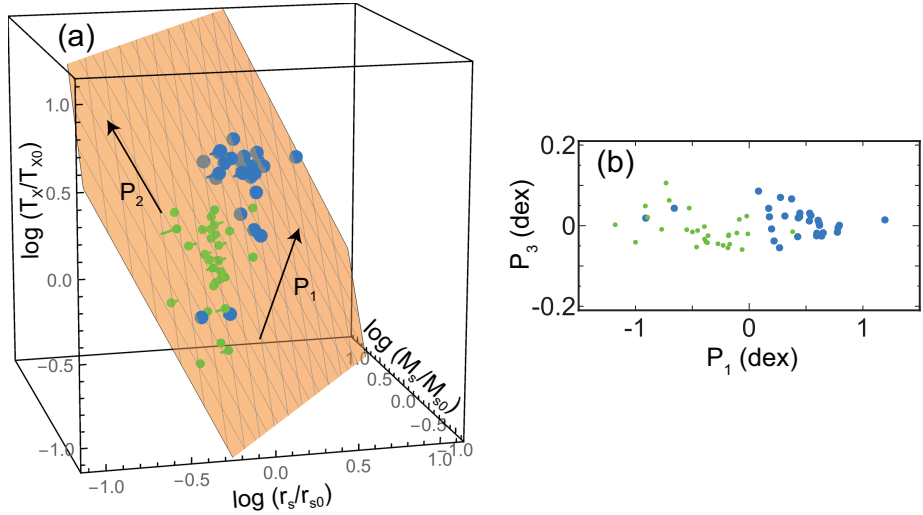} 
%\caption{\hl{The same as
%Figure}~\ref{fig:FP_CLASH} but for the simulations including radiative
%cooling and feedback. The blue (FB0) and the green dots (FB1) are the
%results for $z=0$ and 1, respectively. The axes are normalized by the
%average parameters of the combined sample ($r_{s0}=388$~kpc,
%$M_{s0}=1.4\times 10^{14}\: M_\odot$, and $T_{X0}=4.8$~keV). (Figure is
%reconstructed from Figure~4 of \cite{fuj18a}).\textcolor{red}{We appreciate full descriptions re-written here instead of `the same as Figure 1' .\\please explain the subfigures a and b in the caption. }}  \label{fig:FP_RASIA}
%%please explian the subfigure a and b in the caption.
\caption{(\textbf{a}) The results of simulations including radiative
cooling and feedback. The blue (FB0) and the green dots (FB1) are
the results for $z=0$ and 1, respectively. The axes are
normalized by the average parameters of the combined
sample ($r_{s0}=388$~kpc, $M_{s0}=1.4\times 10^{14}\: M_\odot$,
and $T_{X0}=4.8$~keV).  The orange plane is the best fit of the
data. The arrow $P_1$ shows the direction to which the data
distribution is most extended, and the arrow $P_2$ is
perpendicular to $P_1$ on the plane.
(\textbf{b}) The cross-section of the plane in (\textbf{a}). The
direction $P_3$ is the plane normal
(Figure is reconstructed from Figure~4 of \cite{fuj18a}).}
\label{fig:FP_RASIA}
\end{figure}

{ Figure~\ref{fig:FP_RASIA} presents the results of other numerical
simulations including phenomena such as heating by AGNs and SNe in
addition to radiative cooling (see details in
\cite{ras15a,fuj18a}). Sample FB0 (blue dots) consists of the clusters
at $z=0$, while sample FB1 (green dots) refers to the runs at $z=1$.
These~runs simulate 29 Lagrangian regions around massive clusters with
$M_{\rm 200}\sim 1$--$30\times 10^{14}\: h^{-1}\: M_\odot$ at $z=0$. The~mass resolution for the dark-matter particles and the initial gas
particles are $m_{\rm DM}=8.3~\times~10^8\: h^{-1}\: M_\odot$ and $m_{\rm
SPH}=1.5\times 10^8\: h^{-1}\: M_\odot$, respectively. In
high-resolution regions the gravitational softening is set to be $3.75\:
h^{-1}$~kpc \citep{2017MNRAS.468..531B}. For the gas particles, this is
always in comoving units, while for the DM particles it changes to
physical units from $z=2$ to $z=0$.}  For these samples, the temperature
is estimated for $r=50$--500~kpc, and thus the influence of cool cores
is removed. Both~groups of dots are located on almost the same
fundamental plane, and the plane angles for the two samples are almost
the same (Figure~\ref{fig:P3angle}). This means that clusters evolve
along the unique plane in the direction of $P_1$ in
Figure~\ref{fig:FP_RASIA}a. The plane angles for FB0 and FB1 are not
much different from those for the CLASH data and the MUSIC adiabatic
simulations (Figure~\ref{fig:P3angle}). In Figure~\ref{fig:P3angle}, NF0
is the result of a simulation that is the same as FB0 but not including
radiative cooling and feedback. Since their angles are almost the same,
this indicates that radiative cooling and feedback do not much affect
the fundamental plane. This is because we are discussing cluster
properties on a scale of $r\sim r_s \gtrsim 300$~kpc, and the influences
of cool cores, where radiative cooling and feedback are especially
important, are~ignorable.
\section{Origin of the Fundamental Plane and Cluster Growth}
\label{sec:origin}

Fujita et al. \cite{fuj18a} explained the origin of the fundamental
plane using an analytic similarity solution developed by Bertschinger
\cite{ber85a} (see also \cite{shi16b}). This solution treats spherical
collapse of an overdense region in the Einstein--de Sitter universe
($\Omega_0=1$) and subsequent matter accretion onto the collapsed
object. In the solution, matter profiles are represented by
non-dimensional radius, $\lambda$, density $D(\lambda)$, pressure
$P(\lambda)$, and mass $M(\lambda)$. The solution has a constant called
the ``entropy constant''.
\begin{equation}
\label{eq:enti}
 P(\lambda) D(\lambda)^{-\gamma}
M(\lambda)^{10/3-3\gamma} = {\rm const}\:,
\end{equation}
where $\gamma=5/3$ is the adiabatic index. The non-dimensional
parameters are related to dimensional density $\rho$, pressure $p$, and
mass $m$:
\begin{equation}
\label{eq:nond}
 \rho(r,t) = \rho_{\rm H} D(\lambda)\:,
\hspace{10mm}
p(r,t) = \rho_{\rm H}(r_{\rm ta}/t)^2 P(\lambda)\:,
\hspace{10mm}
m(r,t) = (4\pi/3)\rho_{\rm H} r_{\rm ta}^3 M(\lambda)\:,
\end{equation}
where $r_{\rm ta}(t)$ is the maximum radius that a mass shell reaches
(the turnaround radius), $\rho_{\rm H}\propto t^{-2}$ is the density of
the background universe, and $t$ is the cosmological time. The
non-dimensional radius is given by $\lambda=r/r_{\rm ta}$. {We note
that the similarity solution describes the matter profile in the region
where the matter is later accreted (say, $r\gtrsim r_s$).} The solution
was originally developed for objects totally composed of baryons,  
thus  $\rho$, $p$, and $m$ are for the gas. However, the non-dimensional
profiles ($D$, $p$, and $M$) are not much changed, even if objects are
mostly composed of dark matter \cite{ber85a}. Thus, the profiles $\rho$,
$p$, and $m$ can be regarded as the values for the gas as long as we do
not discuss the normalizations. Although the solution is constructed for
the Einstein--de Sitter universe, it well-reproduces the structure of
objects except for the outermost region even for a $\Lambda$CDM
universe, because the inner region was established when the background
density of the universe was large \cite{ber85a}. From
Equations~(\ref{eq:enti}) and~(\ref{eq:nond}), we obtain
$p\rho^{-5/3}m^{-5/3}\propto A_{\rm ita}^{-3}$, where $A_{\rm
ita}=r_{\rm ita}/t_{\rm ita}^{8/9}$ and is time-independent. Here,
$r_{\rm ita}$ and $t_{\rm ita}$ are the turnaround radius (the maximum
radius before the collapse) and the turnaround time (the time when the
radius reaches turnaround radius) of the overdense region, respectively.
The evolution of the overdense region is described by the conventional
spherical collapse model. Thus, it should follow the spectrum of initial
density perturbations of the universe and the mass of the overdense
region $m_{\rm ita}$ has scaling relations of
\begin{equation}
\label{eq:ita}
 r_{\rm ita}\propto m_{\rm ita}^{(n+5)/6},\hspace{15mm} t_{\rm
ita}\propto m_{\rm ita}^{(n+3)/4}\:,
\end{equation}
where $n$, the spectral index, is of the initial density perturbations
\cite{kai86a} and $n\sim -2$ is expected around the mass scales of
clusters \cite{eis98a,die15a}. Here, we emphasize that
Equation~(\ref{eq:ita}) is applied to the overdense region, and not to
the entire cluster, because we separately treat the initial collapse of
the overdense region and the subsequent matter accretion. Thus, we
obtain $p\rho^{-5/3}m^{-5/3}\propto m_{\rm ita}^{(n-3)/6}$.  Assuming
that $p\propto \rho T_X$, $\rho\propto M_s/r_s^3$ and $m\sim M_s$ at
$r\sim r_s$, it is written as
\begin{equation}
\label{eq:FPita}
r_s^2 M_s^{-7/3} T_X \propto
m_{\rm ita}^{(n-3)/6}\:. 
\end{equation}

The radius $r_{\rm ita}$\footnote{Note that, although the radius $r_{\rm
ita}$ is the turnaround radius of the overdense region, it is
proportional to the radius of the region after the collapse because the
solution is similar.} and the mass $m_{\rm ita}$ of the overdense region
can be connected to the characteristic radius $r_s$ and mass $M_s$ of
the NFW profile. This is because the evolution of both   overdense
region in the similarity solution and the inner region of the NFW
profile ($r\lesssim r_s$) is related to the background universe, and
they evolve in a similar way. In fact, the evolution of the former is
described by the conventional spherical collapse of an overdense region
\cite{ber85a}, and thus the typical density is proportional to that of
the background universe at the collapse. The same applies to the latter
because the characteristic density $\rho_s$ is always $\sim$900 times
as large as that of the background universe at the formation redshift
$z_f$ \cite{cor15c}. Thus, we can assume that $r_s\propto r_{\rm ita}$
and $M_s\propto m_{\rm ita}$, and that the collapse time of the
overdense region ($\sim$2 $t_{\rm ita}$; see,  e.g., \cite{pee80},
section~19) corresponds to the formation redshift $z_f$. { In
summary, the   similarity solution has scales such as $r_{\rm ita}$
and $m_{\rm ita}$. Since $r_{\rm ita}$ and $m_{\rm ita}$ are
proportional to $r_s$ and $M_s$, respectively, the solution has scales
of $r_s$ and $M_s$. These scales represent the border between the
initially-collapsed overdense region and the later-accreted region,
although the actual transition is gradual. This also means that the
initial collapse and the later accretion correspond to the fast-rate
growth phase and the slow-rate growth phase, respectively.}  Finally,
from Equation~(\ref{eq:FPita}), we obtain
\begin{equation}
\label{eq:FP}
 r_s^2 M_s^{-(n+11)/6}T_X={\rm const}\:,
\end{equation}
or $T_X\propto M_s^{(n+11)/6}/r_s^2$. Equation~(\ref{eq:FP}) forms a
plane in the $(\log r_s, \log M_s, \log T_X)$ space. The direction of
the normal when $n=-2$ is shown in Figure~\ref{fig:P3angle} as ``SSol'',
and it is consistent with the CLASH observations and the results of
numerical simulations. Note that this relation~(Equation \eqref{eq:FP})
is independent of redshift $z$ at least $z\lesssim 1$
(Figure~\ref{fig:FP_RASIA}), because $r_s$ and $M_s$ are physical values
that have already reflected the high density of the background universe
in the past.

The similarity solution indicates that clusters are not in virial
equilibrium, because clusters are growing through matter accretion from
their outer environments \cite{ber85a,shi16b}. That is one reason
clusters follow Equation~(\ref{eq:FP}) instead of $T_X\propto M_s/r_s$,
which could be realized if clusters are in virial equilibrium at their
formation. The condition of the virial ``equilibrium'' is represented by
$2\: K + W = 0$, where $K$ is the kinetic and/or thermal energy and $W$
is the gravitational energy. However, according to the virial
``theorem'', {which is a higher-level concept of the virial
``equilibrium''}, additional terms are required when clusters are
growing \cite{ber85a,shi16b}. One is the term representing the increase
of mass and size of clusters and another is the boundary term
originating from the flux of inertia through the boundary and the
pressure at the boundary. The boundary corresponds to the splashback
radius for dark matter and the shock front for gas (see
Section~\ref{sec:intro}).  Note that the similarity solution shows that
clusters are almost in hydrostatic equilibrium in the sense that gas
motion is negligible within clusters even if they are not in virial
equilibrium \cite{ber85a}. The relation between matter accretion and the
cluster structure has also been numerically studied (e.g.,
Ref. \cite{2018arXiv181008212R}).

\begin{figure}[H]
\centering \includegraphics[width=12 cm]{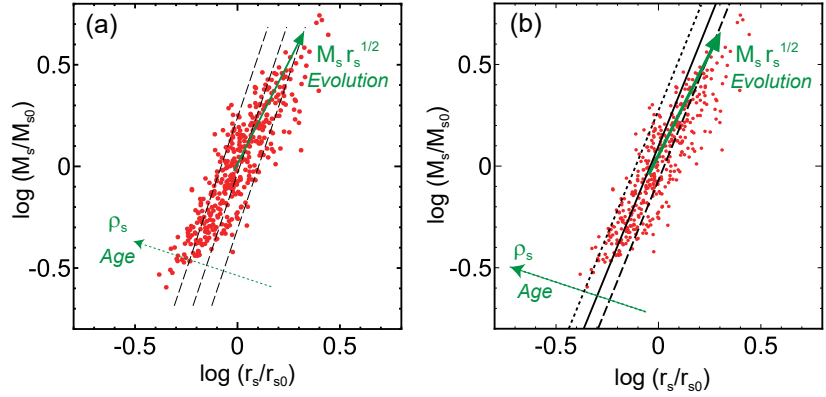} \caption{Projection of
the fundamental plane on the $\log r_s$--$\log M_s$ plane shown in
Figure~\ref{fig:FP_MUSIC}. (\textbf{a})~The red points show the MUSIC clusters
($z=0.25$) and $r_{s0}$ and $M_{s0}$ are the same as those in
Figure~\ref{fig:FP_MUSIC}. The~solid arrow shows the direction of
cluster evolution ($r_s\propto M_s^{1/2}$) and $M_s r_s^{1/2}$ increases
in this direction. The cluster age and $\rho_s$ increase in the
direction of the dotted arrow. Each dashed line satisfies $\rho_s$~=~const
or clusters on a particular line have the same formation redshift
$z_f$. (\textbf{b}) The same as (\textbf{a}) but $M_s$--$r_s$ relation transformed from
$c_\Delta$--$M_\Delta$ relation is drawn (black solid line). Black
dotted and dashed lines correspond to the dispersion of
$c_\Delta$--$M_\Delta$ relation ($\pm$0.1~dex) shown by numerical
simulations (Figures are reconstructed from Figure~5a of \cite{fuj18a}
and Figure~2 of \cite{fuj18b}).}  \label{fig:rM}
\end{figure}

Figure~\ref{fig:rM} shows the projection of the fundamental plane shown
in Figure~\ref{fig:FP_MUSIC} on the $\log r_s$--$\log M_s$ plane. The
solid arrow is parallel to the line of $r_s\propto M_s^{1/2}$ along
which the distribution of the MUSIC clusters (red points) is
elongated. This direction is also close to that of cluster evolution
($P_1$) in Figure~\ref{fig:FP_RASIA} projected on the $\log r_s$--$\log
M_s$ plane \cite{fuj18a}. Since we assumed that $r_s\propto r_{\rm ita}$
and $M_s\propto m_{\rm ita}$, the line corresponds to the first relation
of Equation~(\ref{eq:ita}) when $n=-2$. Considering the derivation of the
relations~in Equation (\ref{eq:ita}) (see \cite{kai86a}), this indicates that the
evolution of clusters on the fundamental plane reflects the spectrum of
the initial density perturbations of the universe and follows
$M_s\propto m_{\rm ita}\propto (1+z_f)^{-6/(n+3)}$
\cite{kai86a}. Figure~\ref{fig:rM}a also shows that the characteristic
density $\rho_s$ decreases as a cluster moves in the direction of the
solid arrow. While the formation redshift $z_f$ is formally related to
the collapse time of the overdense region, in reality, it is often
related to the time of major cluster mergers. That is, the formation
redshift $z_f$ is reset when the cluster experiences a major merger, and
$z_f$ estimated from $\rho_s$ for a given cluster at $z_{\rm obs}$ often
corresponds to the time when the cluster underwent its last major
merger. In fact, numerical simulations have shown that an individual
cluster intermittently moves in the direction of the solid arrow in
Figure~\ref{fig:rM} every time it undergoes mergers \cite{fuj18a}. While
the cluster temporarily deviates the general motion in the middle of a
major merger, the deviation is confined in the fundamental plane and
thus mergers do not much affect the thinness of the plane \cite{fuj18a}.
In other words, the effect of major cluster mergers introduces some
random history that could be different for clusters of the same mass,
but since the mergers move cluster properties within the limits of the
plane, the scatter of the plane does not increase very much.

We would like to point out that in Figure~\ref{fig:rM} simulated
clusters are not tightly distributed along the line of $r_s\propto
M_s^{1/2}$ (solid arrow), and there is a scatter about the line. This
reflects the fact that the density perturbations of the universe are
described by a Gaussian random field (see, e.g., \cite{bar01a}). Thus,
while the variance of the perturbation field $\sigma(M)$ is a decreasing
function of mass scale $M$, the amplitudes of the perturbations
that collapse into objects with a given mass $M$ are not always
$\sigma(M)$. Owing to this, for example, $\rho_s$ and $M_s$ are not
perfectly in one-to-one correspondence, and $\rho_s$ has some range for
a given $M_s$, which produces the band-like distribution of clusters in
Figure~\ref{fig:rM} and on the fundamental plane
(Figures~\ref{fig:FP_CLASH}a, \ref{fig:FP_MUSIC}a and
\ref{fig:FP_RASIA}a). In other words, clusters form a two-parameter
family. Thus, a correlation between two physical quantities is generally
represented by a band rather than a line unless some special combination
of quantities is chosen. In that sense, it is natural that the relation
between $c_\Delta$ and $M_\Delta$ has a large dispersion
\cite{bul01a,lud13a,2008MNRAS.390L..64D,men14a,cor15c}, which will be
discussed in Section~\ref{sec:MT}. On the fundamental plane, different
clusters move along nearly parallel but different tracks each of which
approximately follows the relation of $r_s\propto M_s^{1/2}$
\cite{fuj18a}. While the temperature of each cluster $T_X$ is affected
by its formation time, it also depends on the track the cluster chooses.

\section{Mass--Temperature Relation and the Concentration Parameter}
\label{sec:MT}

The fundamental plane can be used to relate the cluster structure to the
temperature. As an application, we discuss the mass--temperature
relation in this section. It  is well-known that the mass of
clusters and the X-ray temperature has a relation of $M_\Delta\propto
T_\Delta^{3/2}$. This relation is obtained by both observations and
numerical simulations
\cite{bry98a,ett02b,2016A&A...592A...4L,tru18a}. Conventionally, this
relation has been explained based on the following three assumptions:
(i) the typical density of a cluster is $\rho_\Delta =\Delta\rho_c$ (not
$\rho_s$); (ii)~clusters are well-relaxed or virialized, and they are
almost isothermal within $r_\Delta$; and (iii) the X-ray temperature is
determined on a scale of $r_\Delta$ (not $r_s$). Here, we consider
cluster temperature outside cool~cores.

The density $\rho_\Delta$ is represented by $\rho_\Delta\propto \Delta
E(z)^2$, where $E(z)$ is the Hubble parameter at redshift $z$ normalized
by the current value $H_0$. Equation~(\ref{eq:MD}) is associated with
Assumption~(i). From Assumptions~(ii) and (iii), we obtain $T_X\propto
M_\Delta/r_\Delta \propto \rho_\Delta r_\Delta^2 \propto \Delta E(z)^2
r_\Delta^2$. Eliminating $r_\Delta$ by using the relation
$r_\Delta\propto M_\Delta/T_X$, the mass--temperature relation is
obtained:
\begin{equation}
\label{eq:MTclassic}
 M_\Delta \propto T_X^{3/2}\Delta^{-1/2} E(z)^{-1}\:,
\end{equation}
\textls[-20]{which well reproduces the results of observations and simulations
\cite{bry98a,2011ASL.....4..204B,2015SSRv..188...93P}. However, the
assumptions} are clearly inconsistent with the inside-out scenario of
cluster formation and the fundamental plane. For example, the inside-out
scenario indicates that clusters are not well relaxed and keep the
memory of their formation in their structure. The angle of the
fundamental plane shows that clusters are not virialized, as discussed
in Section~\ref{sec:origin}. The NFW profile (Equation~(\ref{eq:NFW}))
is not an isothermal profile ($\rho_{\rm DM}\propto r^{-2}$). These are
inconsistent with Assumption~(ii). Moreover, the tight correlation of
the fundamental plane shows that $T_X$ is determined by $r_s$ and $M_s$,
which contradicts Assumption~(iii).

In \cite{fuj18b}, Fujita et al. showed that the
relation~in Equation (\ref{eq:MTclassic}) can be derived using the fundamental plane
and the halo concentration--mass ($c_\Delta$--$M_\Delta$) relation. The
fundamental plane relation~in Equation (\ref{eq:FP}) is rewritten as
\begin{equation}
\label{eq:TX}
 T_X = T_{X0}\left(\frac{r_s}{r_{s0}}\right)^{-2}
\left(\frac{M_s}{M_{s0}}\right)^{(n+11)/6}\:,
\end{equation}
where $(r_{s0}, M_{s0}, T_{X0})$ corresponds to a representative point
on the fundamental plane, and we adopt $r_{s0}=414$~kpc,
$M_{s0}=1.4\times 10^{14}\: M_\odot$, and $T_{X0}=3.7$~keV based on the
results of the MUSIC simulations~\citep{men14a,fuj18a}. Based on the
inside-out scenario, there are analytical forms of the concentration
parameter represented as a function of $M_\Delta$ and $z$. One example
is
\begin{equation}
\label{eq:duf08a}
 c_{200}(M_{200},z) = 6.71\:\left(\frac{M_{200}}
{2\times 10^{12}h^{-1}M_\odot}\right)^{-0.091}(1+z)^{-0.44}
\end{equation}
for $\Delta=200$ that was obtained by Duffy et al. \cite{2008MNRAS.390L..64D} (see
also
\cite{2013ApJ...766...32B,dut14b,men14a,die15a,2018arXiv180907326D}). From
Equation~(\ref{eq:MD}), we obtain
\begin{equation}
\label{eq:rD}
r_\Delta =
\left(\frac{3 M_\Delta}{4\pi \Delta\:\rho_c(z)}\right)^{1/3}\:.
\end{equation}

Equations~(\ref{eq:cD}), (\ref{eq:duf08a}) and~(\ref{eq:rD}) indicate
that $r_s$ is a function of $M_\Delta$ for a given $z$. Moreover,
Equation~(\ref{eq:MNFW}) suggests that $M_s$ is also a function of
$M_\Delta$:
\begin{equation}
\label{eq:MDMs}
 M_s = M_\Delta\frac{\ln 2-1/2}{\ln(1+c_\Delta)-c_\Delta/(1+c_\Delta)}\:.
\end{equation}

\begin{figure}[H]
\centering \includegraphics[width=12 cm]{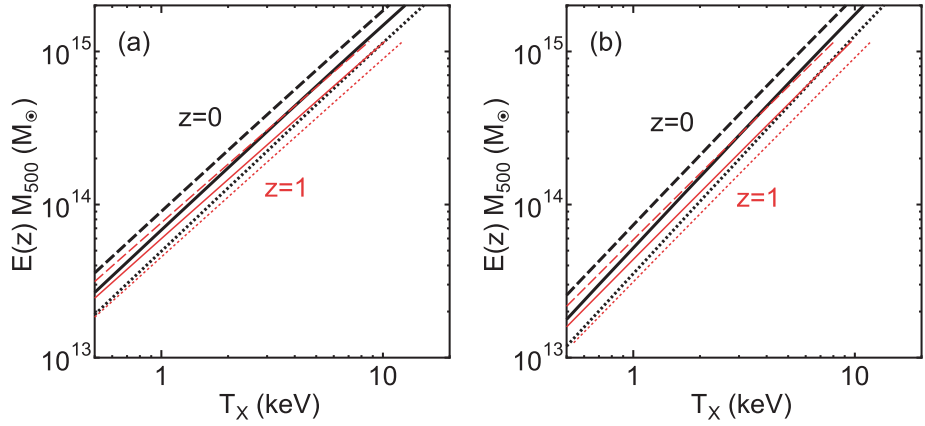}
\caption{$M_\Delta$--$T_X$ relation for $\Delta=500$ derived from the
fundamental plane and the $c_\Delta$--$M_\Delta$ relation (solid
lines): (\textbf{a}) $n=-2$; and (\textbf{b}) $n=-2.5$. The thick black lines and the thin
red lines represent $z=0$ and 1, respectively. Dotted and dashed-lines
correspond to the dispersion of the $c_\Delta$--$M_\Delta$ relation
($\pm$0.1~dex) shown by numerical simulations (Figures are
reconstructed from Figure~1 of \cite{fuj18b}).}  \label{fig:MT}
\end{figure}

Thus, using Equation~(\ref{eq:TX}), $T_X$ can be represented as a
function of $M_\Delta$ for a given $z$. Figure~\ref{fig:MT}a shows the
results for $n=-2$ using a general formula of $c_{200}(M_{200},z)$
developed by Correa et al. \cite{cor15c} instead of
Equation~(\ref{eq:duf08a}).  The slope is $\alpha=1.33$ for $z=0$ and
$\alpha=1.28$ for $z=1$ ($M_\Delta\propto T_X^\alpha$). The slope is
close to but slightly smaller than $\alpha=1.5$. However, the derivation
of the fundamental plane in Section~\ref{sec:origin} might be too
simplified, and there might be some minor uncertainties on $n$
\cite{fuj18b}. In fact, if we take $n=-2.5$, the slope becomes
$\alpha=1.53$ for $z=0$ and $\alpha=1.45$ for $z=1$. Note that, even if
we assume $n=-2.5$, the direction of the fundamental plane
(Equation~(\ref{eq:FP})) is consistent with observations and simulations
\cite{fuj18b}\footnote{Here, we see $n$ as a parameter of the direction
of the fundamental plane, and we do not intend to claim that the
spectral index of the initial density perturbations is exactly
$-$2.5.}. Thus, the relation of $M_\Delta\propto T_X^{3/2}$ can be
reproduced without the virial assumption or $T_X\propto
M_\Delta/r_\Delta$. Note that Figure~\ref{fig:MT} indicates that the red
lines ($z=1$) are slightly below the black lines ($z=0$). This~may cause
some bias about the slope index $\alpha$ if clusters with various
redshifts are plotted at the same time. For example, if higher-redshift
clusters ($z\sim 1$) tend to have smaller masses and lower temperatures
than lower-redshift clusters ($z\sim 0$), the slope is slightly
steepened (larger $\alpha$). We note that
Voit~\cite{2000ApJ...543..113V} (see also~\cite{1998ApJ...500L.111V})
already addressed this issue. He considered accretion history of
clusters and the effects of cluster surfaces as we do. While we focused
on the inner structure of clusters, he studied the evolution of global
properties of clusters.  He concluded that the approximate agreement
between the $M_\Delta$--$T_X$ relation derived via the traditional
collapse model (Equation~(\ref{eq:MTclassic})) and those of simulations
and observations is largely coincidental. Although our approach is
different, our results support the~conclusion.

\begin{figure}[H]
\centering \includegraphics[width=6.5 cm]{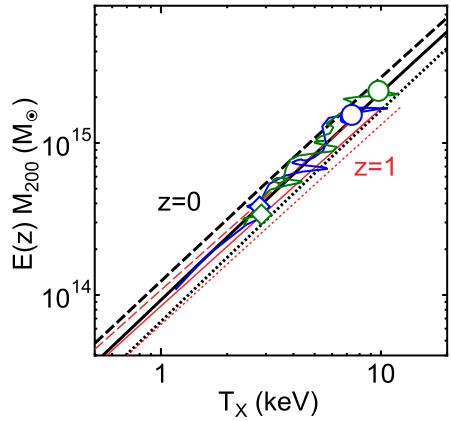} \caption{The same as
Figure~\ref{fig:MT} but   for $\Delta=200$ and $n=-2$. Blue and green
curves present the evolutions of two of the clusters shown in
Figure~\ref{fig:FP_RASIA}. Circles and diamonds show the points of $z=0$
and~1, respectively.}  \label{fig:M200T}
\end{figure}

The relation of $c_\Delta$--$M_\Delta$ or the function
$c_\Delta=c_\Delta(M_\Delta, z)$ can be converted into the relation
between $r_s$ and $M_s$ using Equations~(\ref{eq:MD}), (\ref{eq:cD}),
and (\ref{eq:MDMs}), and the result is shown by the solid black line in
Figure~\ref{fig:rM}b. The black dotted and dashed lines correspond to
the dispersion of $c_\Delta$--$M_\Delta$ relation indicated by numerical
simulations. The three black lines in Figure~\ref{fig:rM}b are almost
parallel to the lines of $\rho_s=$~const or the three black dashed lines
in Figure~\ref{fig:rM}a. This means that the dispersion of
$c_\Delta$--$M_\Delta$ relation is almost the same as that of $\rho_s$
or the dispersion of cluster formation time
$z_f$. Figure~\ref{fig:rM}b also indicates that the minor axis of the
cluster distribution (red points) corresponds to the dispersion of the
$c_\Delta$--$M_\Delta$ relation. The dispersion of the
$c_\Delta$--$M_\Delta$ relation is also associated with that of the
$M_\Delta$--$T_X$ relation, which is indicated by the black dotted and
dashed lines in Figure~\ref{fig:MT}. In Figure~\ref{fig:M200T}, we
present the evolution of simulated clusters along the $M_\Delta$--$T_X$
relation. As   expected, the clusters move along the bands enclosed by
the dotted and dashed lines. The clusters frequently move in the
horizontal direction, which corresponds to temporal temperature increase
during cluster mergers. However, even   during the mergers, the
clusters are located within the bands, which means that the
$M_\Delta$--$T_X$ relation is not much affected by mergers.
\section{Cluster Mass Calibration}

The thinness and solidity of the fundamental plane inspires applications
in cosmology. Here, we show that the plane can be used to calibrate
cluster mass \cite{fuj18b}. Precise estimation of cluster mass is
important. For example, when cosmological parameters are derived from
cluster number counts, scaling relations among observables are used and
they are affected by the calibration of cluster mass~\cite{pla14b}.

\begin{figure}[H]
\centering \includegraphics[width=12 cm]{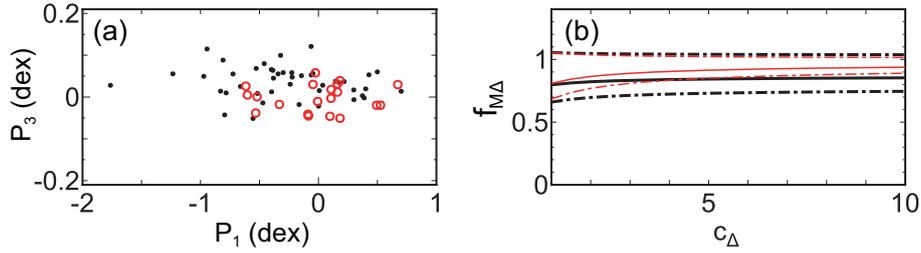} \caption{(\textbf{a})
Cross sections of the fundamental plane. Red open circles are the CLASH
clusters (CFP) and black dots are the X-ray clusters (XFP). (\textbf{b}) Relation
between $f_{M\Delta}$ and concentration parameter $c_\Delta$. Solid
lines are the fiducial relations and the dash-dotted lines show
uncertainties. The difference of black and red lines come from the
different assumptions of the plane shift (see \cite{fuj18b}) (Figures are
reconstructed from Figures~5 and~6 of \cite{fuj18b}).}
\label{fig:cross_bias}
\end{figure}

Figure~\ref{fig:cross_bias}a shows the cross sections of the
fundamental plane. The red open circles are the clusters of the CLASH
sample \cite{pos12a}, for which $r_s$ and $M_s$ are derived through
gravitational lensing. The black dots are those of an X-ray sample
\cite{ett10a}, for which $r_s$ and $M_s$ are derived through X-ray
observations assuming that the ICM is in hydrostatic equilibrium.  We
discuss the fundamental plane formed by the CLASH sample (CFP) and the
one formed by the X-ray sample (XFP) separately. Fixing the direction of
the plane normals at the one shown by SSol in Figure~\ref{fig:P3angle},
the distance between the two fundamental planes is estimated to be
$d_{\rm FP}=0.031^{+0.027}_{-0.039}$~dex in the space of $(\log r_s,
\log M_s, \log T_X)$. Thus, the position of the fundamental planes are
consistent with each other. However, the XFP seems to be located
slightly above the CFP in Figure~\ref{fig:cross_bias}a. The shift
$d_{\rm FP}$ may be caused by a possible systematic difference of
observed $r_s$ or $M_s$ between CFP and XFP because they are obtained
through different methods (gravitational lensing and X-ray
observations). The plane shift in the direction of $r_s$ or $M_s$ can be
estimated from $d_{\rm FP}$. Then, assuming the NFW profile
(Equation~(\ref{eq:NFW}) or (\ref{eq:MNFW})), the shift in the
direction of $M_\Delta$ can be derived \cite{fuj18b}.

Figure~\ref{fig:cross_bias}b shows the systematic difference of
$M_\Delta$, which is defined by $f_{M\Delta}\equiv M_{\Delta
X}/M_{\Delta C}$, where $M_{\Delta X}$ is the mass of a cluster on the
XFP, and $M_{\Delta C}$ is the mass of the same cluster on the
CFP. While the ratio $f_{M\Delta}$ depends on the concentration
parameter $c_\Delta$, the dependence is
weak. Figure~\ref{fig:cross_bias}b shows that $f_{M\Delta} \sim
0.85_{-0.2}^{+0.2}$, which means that the cluster mass estimated through
X-ray observations assuming hydrostatic equilibrium (hydrostatic mass)
is $\sim$15\% smaller than that estimated through gravitational
lensing. Since the error is rather large, the current dataset may not
be accurate enough for the calibration purpose. However, the error could
be reduced by using larger and more accurate datasets in the
future. Assuming that gravitational lensing mass is solid, the value of
$f_{M\Delta} \sim 0.85$ is consistent with the results of numerical
simulations showing that hydrostatic mass tends to be smaller than the
true mass
\cite{nag07a,2008A&A...491...71P,2010A&A...510A..76L,2012NJPh...14e5018R}.

\section{Sparsity}

Finally, we would like to make comments on the halo ``sparsity'', which
has been proposed recently \citep{balmes14,corasaniti18} as a valid
alternative to the full description of the dark matter profile. It
measures the ratio of halo mass at two different radii
(e.g., $M_{500}/M_{1000}$) and, in the case that the halo follows a NFW
profile, it is directly related to the halo concentration. The advantage
in using the halo sparsity is that it has an ensemble average value at a
given redshift with a scatter much smaller than that associated to the
distribution in mass concentration and does not require any modeling of
the mass density profile, which might   significantly deviate from a NFW
one in particular in systems still in the process of complete relaxation,
but only the integrated mass measurements within two overdensities.  The
use of the halo sparsity has   also been proposed as new cosmological
probe for galaxy clusters \citep{corasaniti18} because it carries
cosmological information encoded in the halo mass profile and, at given
redshift, the average sparsity can be predicted from prior knowledge of
the halo mass function.

Both the fundamental plane and the halo sparsity reflect the halo
concentration of clusters. While the fundamental plane gives us the
direct information of cluster formation time, it is generally difficult
to measure $r_s$ and $M_s$ observationally, compared with the
sparsity. In   future study, we will discuss the relation between the
fundamental plane and the halo sparsity.

\section{Conclusions}

It has been known that the concentration of dark halos reflects their
formation history. In particular, the halo structure represented by the
characteristic radius $r_s$, and mass $M_s$ is related to the formation
time of the halo. In this study, we showed that $r_s$, $M_s$, and the X-ray
temperature $T_X$ of observed clusters form a plane (fundamental plane)
in the space of $(\log r_s, \log M_s, \log T_X)$ with a very small
orthogonal scatter. The tight correlation shows that $T_X$ is also
affected by the formation time of individual clusters. Numerical
simulations supported the results and showed that clusters evolve along the
plane. The plane and its angle in the space of $(\log r_s, \log M_s,
\log T_X)$ can be explained by a similarity solution, which indicates
that clusters are still growing and have not reached a state of virial
equilibrium. {In other words, when cluster formation and the
internal structure is considered, matter accretion after the initial
collapse cannot be ignored.} The motion of clusters on the plane was
determined by the spectrum of the initial density perturbations of the
universe. The spread of clusters on the fundamental plane is related to
the scatter of the halo concentration--mass relation.

We also discussed applications of the fundamental plane. For example, we
showed that the mass--temperature relation of clusters ($M_\Delta\propto
T_X^{3/2}$) can be explained by the fundamental plane and the halo
concentration--mass relation without assuming virial equilibrium. We
also showed that the solidity and thinness of the fundamental plane can be
used to calibrate cluster mass. Since the fundamental plane associates
the structure of dark halos with the gas temperature, other applications
may be possible. For example, the gas temperature $T_X$ of a dark halo
could be estimated from $r_s$ and $M_s$ obtained through $N$-body
simulations without calculating gas dynamics.

%%%%%%%%%%%%%%%%%%%%%%%%%%%%%%%%%%%%%%%%%%
\vspace{6pt} 

%%%%%%%%%%%%%%%%%%%%%%%%%%%%%%%%%%%%%%%%%%
%% optional
%\supplementary{The following are available online at \linksupplementary{s1}, Figure S1: title, Table S1: title, Video S1: title.}

% \supplementary{The following are available at \linksupplementary, Figure S1: title, Table S1: title, Video S1: title. A supporting video article is available at doi: link.}

%%%%%%%%%%%%%%%%%%%%%%%%%%%%%%%%%%%%%%%%%%
\authorcontributions{Y.F. coordinated the research, wrote the
 manuscript, obtained the fundamental plane, and made a theoretical
 interpretation. S.E. provided the X-ray data and contributed in the
 interpretation of the findings. K.U. analyzed the data of gravitational
 lensing and contributed in the interpretation of the findings. E.R. and
 M.M. analyzed the numerical simulations, kindly provided by the
 computational groups in Trieste and Madrid, and fit the data with the
 NFW profile. M.D., E.M., N.O., and M.P. contributed in the
 interpretation of the optical and X-ray results and provided extensive
 feedback on the study.}

%%%%%%%%%%%%%%%%%%%%%%%%%%%%%%%%%%%%%%%%%%
\funding{This work was supported by MEXT KAKENHI No.~18K03647
(YF). S.E. acknowledges financial contribution from the contracts NARO15
ASI-INAF I/037/12/0, ASI 2015-046-R.0 and ASI-INAF n.2017-14-H.0.
K.U. acknowledges support from the Ministry of Science and Technology of
Taiwan (grant MoST 106-2628- M-001-003-MY3) and from Academia Sinica
(grant AS-IA-107-M01). E.R. acknowledge support from the ExaNeSt and
EuroExa projects, funded by the European Union's Horizon 2020 research
and innovation programme under grant agreements No. 671553 and No. 754337,
respectively.}

%%%%%%%%%%%%%%%%%%%%%%%%%%%%%%%%%%%%%%%%%%
\acknowledgments{We thank referees for their useful comments, which were
very helpful to clarify explanations.}

%%%%%%%%%%%%%%%%%%%%%%%%%%%%%%%%%%%%%%%%%%
\conflictsofinterest{The authors declare no conflict of interest.}

\end{document}